# Localized plasmonic meron-antimeron pairs in doubly degenerate orbitals

Jie Yang[1†], Xinmin Fu[1†], Jiafu Wang[1,2,3*], Yifan Li[4], Jingxian Zhang[1], Fangyuan Qi[1], Yajuan Han[1], Yuxiang Jia[1], Guy A E Vandenbosch[4], Tie Jun Cui[2,5*], and Xuezhi Zheng[4,6‡]

[1]*Shaanxi Key Laboratory of Artificially-Structured Functional Materials and Devices, Xi'an 710051, China*
[2]*Suzhou Laboratory, Suzhou 215123, China*
[3]*Xi'an Jiaotong University, Xianning West Road 28, Xi'an, 710049, China.*
[4]*WaveCoRE research group, KU Leuven, Kasteelpark Arenberg 10, BUS 2444, Leuven B-3001, Belgium*
[5]*State Key Laboratory of Millimeter Wave, Southeast University, Nanjing 210096, China*
[6]*Center for Polariton-driven Light-Matter Interactions, University of Southern Denmark, Campusvej 55, Odense 5230, Denmark*
\*E-mail: wangjiafu1981@126.com; ‡E-mail: tjcui@seu.edu.cn; \*E-mail: xuezhi.zheng@esat.kuleuven.be
†The authors contribute equally.

**Abstract** Topological defects are pivotal in elucidating kaleidoscopic topological phenomena in different physical systems. Meron-antimeron pairs are a type of topological defects firstly found as soliton solutions to SU(2) Yang-Mills equations in gauge theory, and then identified in condensed matter physics as a type of magnetic quasiparticles created in the context of topological charge conservation. Here, we show that isolated meron-antimeron pairs constitute a new form of optical topological quasiparticles that naturally emerge in doubly degenerate orbitals of plasmonic systems, including fundamental and higher-order ones, and their target-type counterparts. We demonstrate that their topological charges are strictly imposed by orbital indices from the doubly degenerate irreducible representations (irreps) of groups consisting of rotational symmetries, and thus are upper-bounded by the orbital indices imposed by group theory. In addition, we find that there exist highly-localized isolated (anti)merons in plasmonic spin textures, which were previously observed mostly in the form of lattices or clusters. We further demonstrate a locking effect between the chirality of the (anti)merons and the parity of the irreps. Then, the topological origins of the revealed topological quasiparticles, i.e., phase, V-point and L-line singularities in plasmonic fields, are investigated. Finally, a complete symmetry classification of the topological quasiparticles is provided. Generalizing the meron-antimeron pairs to photonic systems provides various possibilities for the applications in optical vectorial imaging, deep-subwavelength sensing and metrology.

**Keywords**: Optical quasiparticle; meron-antimeron pair; localized surface plasmon; doubly degenerate orbital; symmetry; spin texture

## Introduction

As topological defects, merons were firstly found as soliton solutions to Yang-Mills equations in gauge theory in 1976[1]. Thereafter, they are generalized to and identified in condensed matter physics as a type of magnetic quasiparticles topologically equivalent to one-half of a magnetic skyrmion[2], further inspiring many spintronics applications like information storage and transfer[3] or constructing qubits for quantum computation[4] owing to their topological stability and peculiar vectorial configurations even at small sizes. Recently, motivated by the advances of magnetic quasiparticles and largely fueled by the explosive development of singular optics and topological photonics, optical analogues to magnetic quasiparticles, namely, using various electromagnetic (EM) field vectors to emulate the vector configurations of magnetic quasiparticles, have been attracting growing attention and rapidly become a hot topic in optics and photonics[5–7]. Several forms of optical quasiparticles have been intensely studied, especially optical merons, skyrmions and their transformations or extensions[8,9]. These optical quasiparticles exhibit great topologically protected stability and deep-subwavelength features, enable new degrees of freedom to structure light and promise many singularity-based applications[10–12]. Nevertheless, there are still many other forms of optical quasiparticles remaining to be further explored[8,13]

Meron pairs are two-meron configurations originating from pairing of a meron and itself or its antiparticle, antimeron, which are soliton solutions to SU(2) Yang-Mills equations and may lead to quark confinement in quantum chromodynamics[14–17]. As a generic topological defect in wave fields, they have also been identified in condensed matter physics, including two categories, (anti)meron-(anti)meron pairs and meron-antimeron pairs[14,18]. The former mostly known as (anti)bimerons carries a net topological invariant of 1(-1) and thus is topologically equivalent to a (anti)skyrmion[19,20]. The latter carries a net topological invariant of 0 and is created in the context of topological invariant conservation[18,21], similar to the Schwinger mechanism, creating particle-antiparticle pairs[22] or monopole-antimonopole pairs[23]. As the most fundamental localized topological configuration, the meron-antimeron pairs can be applied to construct meron lattices or networks and inspire many applications in spintronics[18,24,25]. In optics, the isolated (anti)meron-(anti)meron pairs have been explored mostly from the aspect of bimerons[26–28], but the isolated optical meron-antimeron pairs have so far remained untapped, which until now have been only observed in the form of meron lattices[29,30].

Here, from a pure symmetry consideration we find a new form of isolated meron-antimeron pairs carrying orbital angular momentum (OAM) in doubly degenerate orbitals of plasmonic systems with rotational symmetries, including fundamental and higher-order ones, and their target-type counterparts[31]. They are created in plasmonic electric and magnetic field components of localized surface plasmons (LSPs)[32], and thus are harmonic and highly-localized. Their topological charges, different from the well-known skyrmion number, are characterized by a so-called absolute skyrmion number, which also represents the meron-antimeron pair numbers in the vectorial configurations. We demonstrate that the absolute skyrmion numbers of the pairs are strictly imposed by orbital indices of doubly degenerate irreducible representations (irreps) of the groups composing of rotational symmetries (also known as irrep indices)[33]. The absolute skyrmion numbers are upper bounded since the irrep indices are limited by the nearest-lower integer of $(M-1)/2$, i.e., $\lfloor \frac{M-1}{2} \rfloor$, where $M$ is the rotational symmetry dimensions[33]. Furthermore, we find that in plasmonic spin textures there exist highly localized isolated (anti)merons, which are mostly observed in the form of lattices or clusters[5,18,26]. We further demonstrate that the chirality of the (anti)meron is locked with the parity of the irrep. In addition, by extending our discussions to nondegenerate irreps, e.g., $B_1$ irrep in the case of even $M$, a steady spin meron-antimeron pair and $(M/2)$-order harmonic field meron-antimeron pair are discovered, both of which carry no OAM due to the nondegeneracy of the irrep. Then, the topological origins of the above optical quasiparticles are elaborated, i.e., the phase, V-point and L-line singularities in plasmonic fields. All revealed optical quasiparticles exhibit deep-subwavelength feature down to $\lambda/18$. Experiments well demonstrate our results. Finally, a complete symmetry classification of the topological quasiparticles is provided. Our findings shed light on the meron-antimeron pairs in optics and can be extended to generic wave systems owing to the generality of the symmetry arguments.

## Meron-antimeron pairs, merons and antimerons in plasmonic fields

Ring resonators made of metal or dielectric are typical photonic microstructures with rotational symmetry, showing great importance in studying plasmonics, singular optics and non-hermitian photonics[32,34–37]. Here, we design a ring resonator made of metal to validate our findings (see details in Fig. 1a and SI 1). The resonator is known to support the spoof LSPs at microwave or terahertz frequency[32,34]. It holds 8-fold rotational and reflection symmetries. These symmetries form an 8-order dihedral group, i.e., $D_8$ group (see details in SI 2)[33]. According to the interplay theory of symmetry and light-microstructure interaction, eigenstates of the resonator can be categorized by the irreducible representations (irreps) of $D_8$ group[38].

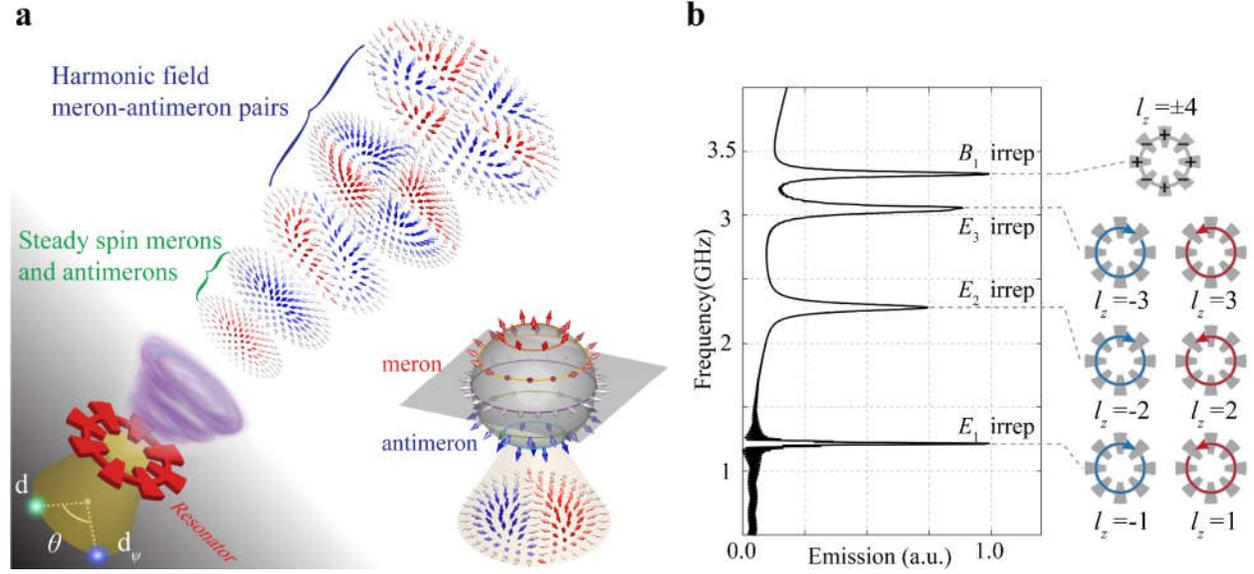

**Fig. 1. Emission of isolated merons, anti-merons and meron-antimeron pairs. a,** The illustration of plasmonic merons, antimerons and meron-antimeron pairs emitted from the ring resonator, which is excited by two point sources marked by green and blue glowing dots. The right-bottom panel illustrates the mapping from a (anti)meron to the (lower)upper half unit sphere. **b,** The emission spectrum of the resonator excited by a point source, where four irreps are excited at 1.22, 2.28, 3.06, and 3.32 GHz, corresponding to $E_1, E_2, E_3$, and $B_1$ irreps, respectively. Of these, the $E_1, E_2,$ and $E_3$ irreps correspond to three different double degeneracies with OAMs of $\pm 1, \pm 2,$ and $\pm 3$, respectively; and the $B_1$ irrep is nondegenerate, whose basis function (or eigenstates) can be seen as a superposition of states with orbits of $\pm 4$. The blue or red directed circle denotes the parity of OAM of -1 or 1 carried by one of the doubly degenerate states, which is totally defined by the symmetry of each dimension of an irrep (see SI 2).

The group defines seven irreps, including four nondegenerate (marked by $A_1, A_2, B_1,$ and $B_2$) and three doubly degenerate (marked by $E_1, E_2,$ and $E_3$) resulting from the reflection or time-reversal symmetry[33]. Therefore, the eigenstates of the resonator can be classified into seven categories, including four singlet (i.e., nondegenerate) and three doublet (i.e., doubly degenerate) states (see details in SI 2). Each eigenstate can be excited by an incident field if and only if a so-called symmetry-matching condition is satisfied, i.e., the incident field has nonvanishing projection along the irrep to which the eigenstate belongs[38]. Such an incident field can be customized by various EM fields, e.g., Laguerre-Gaussian modes[39] or radiated EM waves of several point sources like molecular chromophores[40] in visible light or infinitesimal dipoles[41] at microwaves or terahertz.

Next, by employing symmetry principles we customize the radiated EM waves of several point sources to excite the resonator.

We firstly consider one point-source configuration. Fig. 1b depicts the emission spectrum of the resonator simulated by CST Microwave Studio (see details in SI 1), from which we observe that four eigenstates are excited at four frequencies, corresponding to $E_1$, $E_2$, $E_3$, and $B_1$ irreps, respectively. Of these, the eigenstates in the $E_1$, $E_2$, and $E_3$ irreps are doublet, which can exhibit vorticity and carry OAMs of $\pm\hbar$, $\pm 2\hbar$, and $\pm 3\hbar$, respectively, owing to the double degeneracy of these irreps (note that from here the reduced Plank constant $\hbar$ will be suppressed). The eigenstate in the $B_1$ irrep is singlet, which exhibits no vorticity because of nondegeneracy of the irrep. Here we focus on the eigenstates in doubly degenerate orbitals and the $B_1$ irrep will be discussed in the last section and in SI 1.

To break the double degeneracies and selectively excite one of the doublet states, we need to use at least two point sources to excite the resonator instead of one[42]. The two point sources are marked by d and $d_\psi$ in Fig. 1a, which are linked with a rotation of $\theta$ with respect to the rotationally symmetric axis and differ with a dynamic phase of $\psi$. Assume that the electric field radiated by the point source d is $\mathbf{E}_d(\mathbf{r})$. Then the total electric fields radiated by the two point sources can be cast as $\mathbf{E}_d(\mathbf{r}) + e^{i\psi}\hat{P}_\theta\{\mathbf{E}_d(\mathbf{r})\}$, where $\hat{P}_\theta$ denotes the rotation of $\theta$. By group theory arguments, we introduce the following constraint conditions for $\theta$ and $\psi$,[42]

$$\theta = \frac{\pi}{2l_z} + (n_2 - n_1)\frac{\pi}{l_z}, \quad \psi = \frac{\pi}{2} + (n_1 + n_2)\pi, \tag{1}$$

where $n_1$ and $n_2$ are arbitrary integers. If and only if the above conditions are satisfied, then the eigenstate carrying OAM of $l_z$ will be exclusively excited and the one carrying OAM of $-l_z$ will be suppressed, namely, the doubly degenerate orbital in the eigenstate will be lifted. Based on Eq. (1), we design several two-port feeding networks to excite the resonator so that different doubly degenerate eigenstates belonging to the $E_1$, $E_2$, and $E_3$ irreps can be selectively excited (see SI 1). In these doubly degenerate orbitals, many types of topological quasiparticles like (target)meron-antimeron pairs and (anti)merons surprisingly emerge from localized plasmonic (magnetic)electric and spin fields.

**Harmonic meron-antimeron pairs in plasmonic electric fields**

Fig. 2 depicts the magnitude distributions and vectorial configurations of normalized plasmonic electric fields $\mathbf{e}$, where $\mathbf{e} = \text{Re}\{E_x, E_y, E_z\}/|\mathbf{E}|$ and $\mathbf{E}$ denotes the localized plasmonic electric fields. From the top panel of Fig. 2a, it can be observed that the excited eigenstate exhibits a dumbbell magnitude distribution and its vectorial configuration is a combination of typical Néel-type meron and antimeron with skyrmion numbers of 1/2 and -1/2, respectively[3]. The skyrmion number of the (anti)meron is evaluated by $\frac{1}{4\pi}\iint D dx dy$, where $D = \mathbf{e} \cdot (\partial_x \mathbf{e} \times \partial_y \mathbf{e})$ is the skyrmion number density[6]. Such a vectorial configuration defines the meron-antimeron pair. Since the skyrmion numbers of meron and antimeron are opposite, the net skyrmion number of the meron-antimeron pair is zero, thus inadequate to characterize the topological vectorial configuration of the pair[14,18]. Here we follow the convention in condensed matter physics and adopt an absolute skyrmion density $D_{\text{abs}} = \sum_i |D_i|$ to evaluate their topological charges[21], where $D_i$ is the skyrmion number density of meron ($i=1$) or antimeron ($i=2$) in the pair. Then, the absolute skyrmion number of the meron-antimeron pair can be defined as[21],

$$N = \frac{1}{4\pi}\iint_A D_{\text{abs}} dx dy = |l_z|, \qquad (2)$$

where $A$ denotes the integration region and again, $l_z$ denotes the orbital/irrep index. The absolute skyrmion number also represents the meron-antimeron pair number in the vectorial configuration. With Eq. (2), the absolute skyrmion number of the meron-antimeron pair can be evaluated as 1. From the bottom panel of Fig. 2a, a similar meron-antimeron pair is observed as well, whose absolute skyrmion number is 1. The two meron-antimeron pairs are dynamically rotated in the opposite directions and carry opposite OAMs of -1 and 1, respectively (see Fig. 5a). Therefore, the doubly degenerate orbital of the $E_1$ irrep corresponds to two meron-antimeron pair states with opposite OAMs of $\pm 1$. For the $E_2$ irrep, we observe from Fig. 2b that both of the excited eigenstates exhibit a quadrupole magnitude distribution and their vectorial configurations are a combination of two pairs of Néel-type merons and antimerons, thus defining a second-order meron-antimeron pair with the absolute skyrmion number of 2. The two second-order meron-antimeron pairs are rotated in the opposite directions and carry opposite OAMs of -2 and 2, respectively (see SI 3). For the $E_3$ irrep, two third-order Néel-type meron-antimeron pairs with the absolute skyrmion numbers of 3 are observed in Fig. 2c, which also carry opposite OAMs of -3 and 3, respectively (see SI 3). Therefore, here we have two important observations: 1) the absolute skyrmion numbers of meron-antimeron pair states are locked with the irrep indices of the group; 2) OAM of each meron-antimeron pair is locked with parity of the doubly degenerate orbits of the irrep and accordingly, can be fully controlled by the handedness of the external excitation. It is worth noting that the skyrmion numbers of meron or antimeron in each pair state are averaged to zero over one optical oscillation cycle, consistent with the reported works about the skyrmions/merons composed of plasmonic fields[6,12,43]. Hence, we refer the meron-antimeron pairs to harmonic ones.

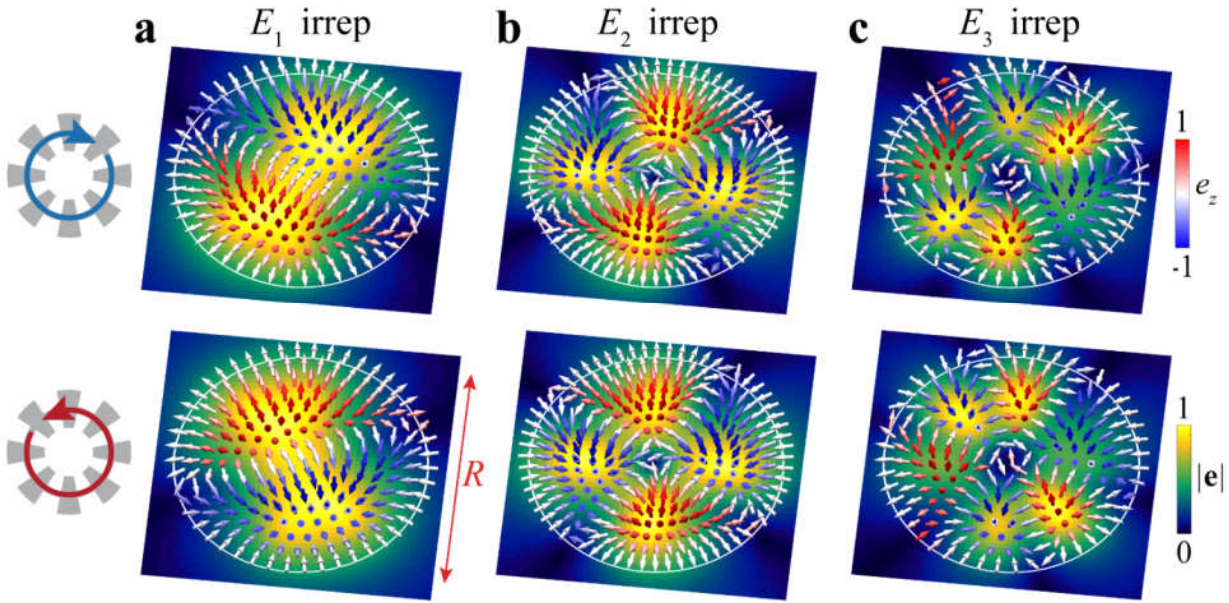

**Fig. 2. Harmonic field meron-antimeron pairs in the plasmonic electric field vectors. a,** Two excited meron-antimeron pairs with absolute skyrmion numbers of 1 but opposite OAMs of -1 and 1 at 1.19 GHz, corresponding to the $E_1$ irrep. **b,** Two excited meron-antimeron pairs with absolute skyrmion numbers of 2 but opposite OAMs of -2 and 2 at 2.25 GHz, corresponding to the $E_2$ irrep. **c,** Two excited meron-antimeron pairs with absolute skyrmion number of 3 but opposite OAMs of -3 and 3 at 2.99 GHz, corresponding to the $E_3$ irrep. $|e|$ denotes the normalized magnitude distributions of electric fields, and $e_z$ denotes the z component of normalized electric fields. The colors of the vectors are encoded from blue to red to denote the latitude angle $e_z$ varying from -1 to 1. The diameter of the white circles is $R = 48mm$. In **a**, the meron-antimeron pairs are on the deep-subwavelength scale down to $\lambda/5.3$, indicating the scale of each meron/antimeron in the pair down to $\lambda/10.6$.

**Harmonic target meron-antimeron pairs in plasmonic magnetic fields.**

Apart from electric fields, harmonic meron-antimeron pairs also emerge from the magnetic fields of spoof LSPs, where only skyrmions were believed to exist in the previous study[44]. Optical (anti)merons in the pairs even exhibit topological configurations of so-called target skyrmions, namely, their vector distributions flip an angle of $\phi$ larger than $\pi$ along the radial direction[8,31]. For example, the vectors of the meron or antimeron in Fig. 2a only flip an angle of $\pi$ along the radial direction and thus are not target ones.

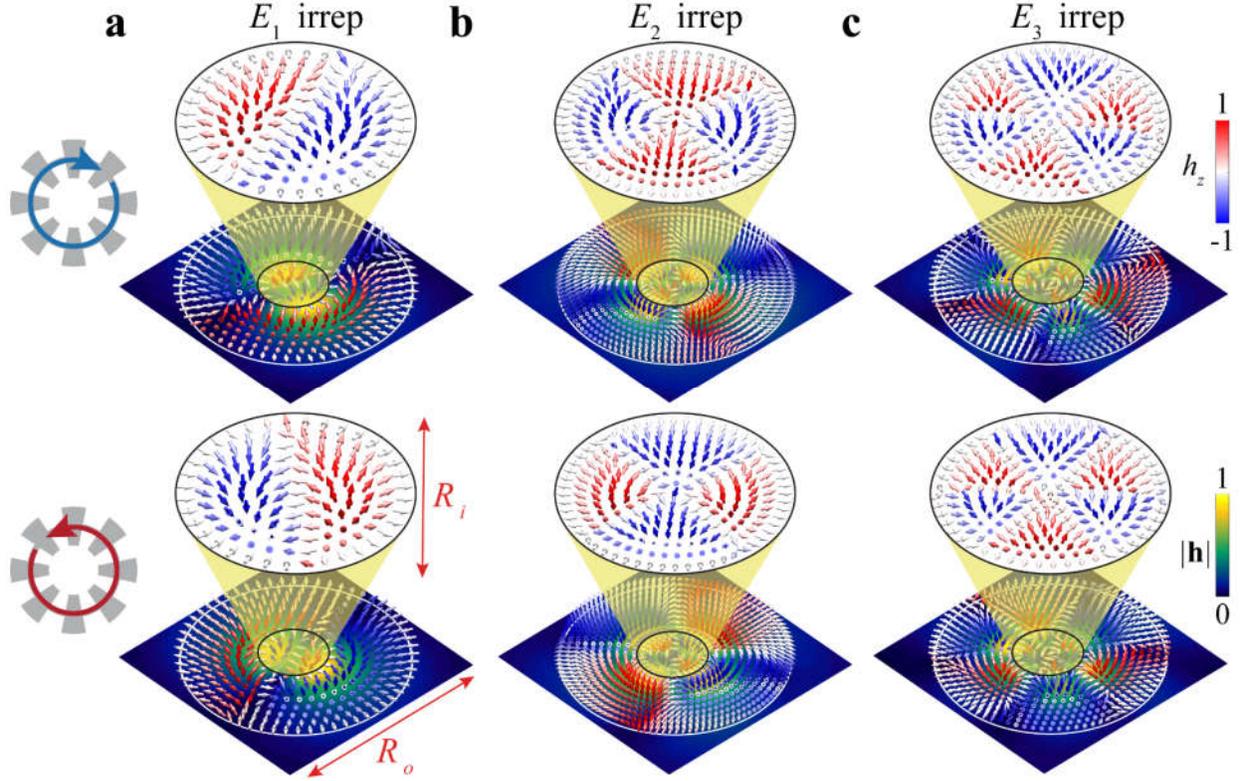

**Fig. 3. Harmonic target meron-antimeron pairs in plasmonic magnetic field vectors. a,** Two excited meron-antimeron pairs with the absolute skyrmion number of 1 but opposite OAMs of -1 and 1 at 1.19 GHz, corresponding to two dimensions of the $E_1$ irrep. **b,** Two excited meron-antimeron pairs with the absolute skyrmion number of 2 but opposite OAMs of -2 and 2 at 2.25 GHz, corresponding to two dimensions of the $E_2$ irrep. **c,** Two excited meron-antimeron pairs with the absolute skyrmion number of 3 but opposite OAMs of -3 and 3 at 2.99 GHz, corresponding to two dimensions of the $E_3$ irrep. $|\mathbf{h}|$ denotes the normalized magnitude distributions of magnetic fields, and $h_z$ denotes the $z$ component of normalized magnetic fields. The colors of the vectors are encoded from blue to red to denote the latitude angle $h_z$ varying from -1 to 1. The inner and outer diameters of the circles denoted by black and white colors are $R_i = 28 mm$ and $R_o = 88 mm$, respectively. The meron-antimeron pairs in the inner circular regions are further presented in the zoomed-in panels. In **a**, the inner meron-antimeron pairs are on the deep-subwavelength scale down to $\lambda/9$, indicating the scale of each meron/antimeron in the pair down to $\lambda/18$.

Fig. 3 depicts the vectorial configurations of normalized plasmonic magnetic fields **h** of three doublet states, where $\mathbf{h} = \text{Re}\{H_x, H_y, H_z\}/|\mathbf{H}|$ and **H** denotes the localized plasmonic magnetic fields. For the $E_1$ irrep, from the top panel of Fig. 3a we observe that 1) a meron-antimeron pair with the absolute skyrmion number of 1 occurs at the zoomed-in inner circular region encircled by a black solid circle; and 2) meron (antimeron) in the pair is surrounded by another antimeron (meron) in the outer annular region encircled by the black and white solid circles, i.e., in the outer annular region another meron-antimeron pair with the absolute skyrmion number of 1 is formed. Therefore, in the whole region encircled by the solid black circle, the magnetic field vectors flip an angle of $2\pi$ along the radial direction and thus the meron-antimeron pair in Fig. 3a is a target one. From the bottom panel of Fig. 3a, a similar target meron-antimeron pair is observed. Note that the two pair states are doubly orbital degenerate and carry opposite OAMs of -1 and 1, respectively (see Fig. 5c). For the $E_2$ irrep, from Fig. 3b we observe that two target meron-antimeron pairs with the absolute skyrmion number of 2 are formed, carrying opposite OAMs of -2 and 2, respectively (see SI 3). For the $E_3$ irrep, from Fig. 3c we observe that two target meron-antimeron pairs with the absolute skyrmion number of 3 are formed, carrying opposite OAMs of -3 and 3, respectively (see SI 3). This magnetic-field case also demonstrates our previous observations, i.e., the absolute skyrmion numbers are locked with the doubly degenerate irrep indices; and the OAM of each meron-antimeron pair is locked with parity of the doubly degenerate orbits of the irrep. It is worth noting that the target meron-antimeron pairs shown here are harmonic due to the time periodicity of plasmonic magnetic fields.

**Steady isolated (anti)merons in plasmonic spin textures**

Different from the harmonic meron-antimeron pairs in the plasmonic electric/magnetic fields, we find that the time-invariant (i.e., steady) isolated (anti)merons emerge from plasmonic spin textures, which normally exist only in pairs or groups in continuous geometries[18]. The spin vector **S** of spoof LSPs is defined as, $\mathbf{S} = \frac{1}{2\omega}\text{Im}(\varepsilon\mathbf{E}^* \times \mathbf{E} + \mu\mathbf{H}^* \times \mathbf{H})$, where $\omega$ is the angular frequency; $\varepsilon$ and $\mu$ are the permittivity and permeability, respectively; and the asterisks denote complex conjugate. Accordingly, the normalized spin vector **s** can be cast in the form of $\mathbf{s} = \mathbf{S}/|\mathbf{S}|$. Fig. 4 depicts the normalized spin textures of three doublet states. From the top(bottom) panels of Fig. 4a-c, we observe that antimerons(merons) with the skyrmion numbers of $-\frac{1}{2}(\frac{1}{2})$ appear in the spin textures belonging to the first(second) dimension of the $E_1$, $E_2$, and $E_3$ irreps, and the related spin intensity and energy flux are highly localized in deep-subwavelength regions. Therefore, the parity of (anti)merons (denoted with $C$) is completely locked with the parity of the doubly degenerate orbitals, i.e.,

$$C = \text{sgn}(l_z), \tag{3}$$

and accordingly, can be fully controlled by the handedness of the external excitation. The topologically nontrivial vectorial configurations of the spin (anti)merons in Fig. 4 are originated from the ones of the meron-antimeron pairs in the plasmonic electric/magnetic fields in Fig. 2 and Fig. 3. Their formation can be explained in the framework of plasmonic SOI[7,45,46]. It is worth noting that in Fig. 4b-c, the normalized spin vectors at the central points are singular since the spin intensities at the centers are zero. Hence, the (anti)merons corresponding to the higher-index irrep like $E_2$ and $E_3$ do not have very spin-up (spin-down) vectors at the centers, which actually can be defined by their neighboring spin vectors just like the case of defining the higher-order polarization singularity than $\pm 1/2$.[47] The steady isolated meron-antimeron pair also exists in the plasmonic spin textures, which will be discussed later on the $B_1$ irrep.

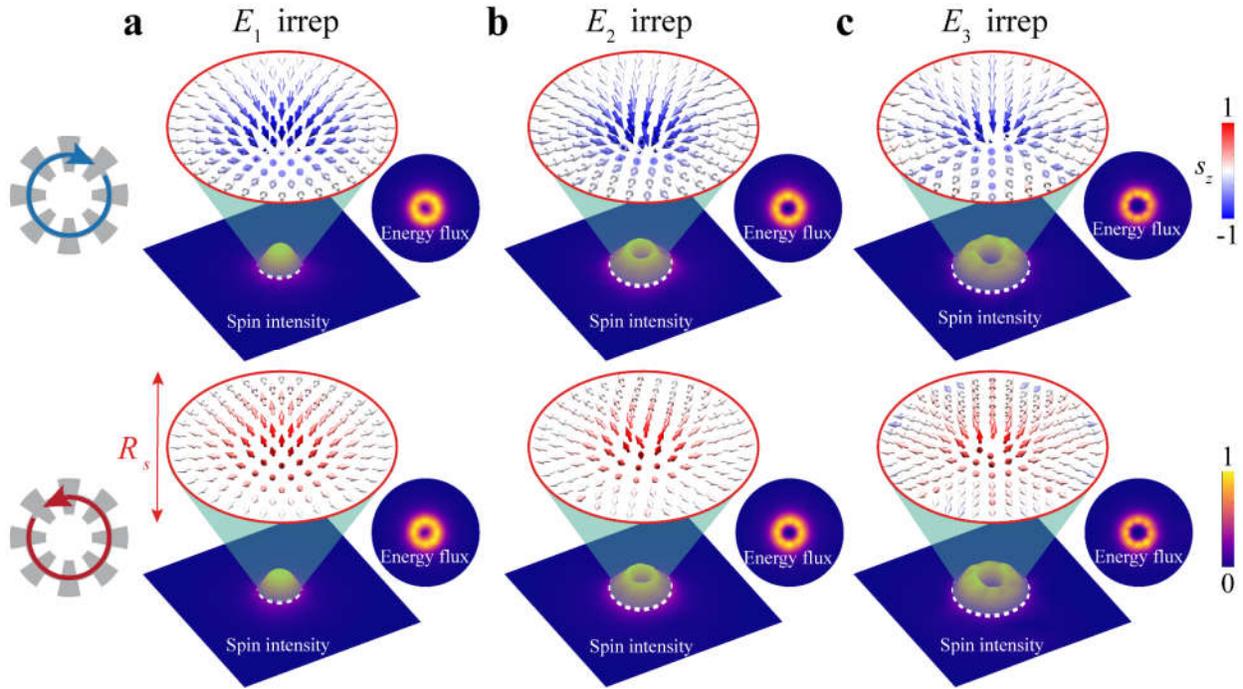

**Fig. 4. Steady merons and antimerons in plasmonic spin textures. a,** Meron and antimeron with opposite OAMs of -1 and 1 at 1.19 GHz, corresponding to two dimensions of the $E_1$ irrep. **b,** Meron and antimeron with opposite OAMs of -2 and 2 at 2.25 GHz, corresponding to two dimensions of the $E_2$ irrep. **c,** Meron and antimeron with opposite OAMs of -3 and 3 at 2.99 GHz, corresponding to two dimensions of the $E_3$ irrep. $s_z$ denotes the $z$ component of normalized spin vectors. The colors of the vectors are encoded from blue to red to denote the latitude angle $s_z$ varying from -1 to 1. The diameter of the red dashed circles is $R_s = 28mm$. In the circular inset of each figure, the magnitude distributions of energy flux $\boldsymbol{p} = \frac{1}{2}Re(\boldsymbol{E} \times \boldsymbol{H}^*)$ are plotted, from which we observe that the energy is highly localized in the subwavelength regimes. In **a**, the meron/antimeron is on the deep subwavelength scale down to $\lambda/9$. In each plot of spin intensity, spin vectors in the area without spin vectors plotted either are in-plane oriented or vanish since the spin intensity is zero over there.

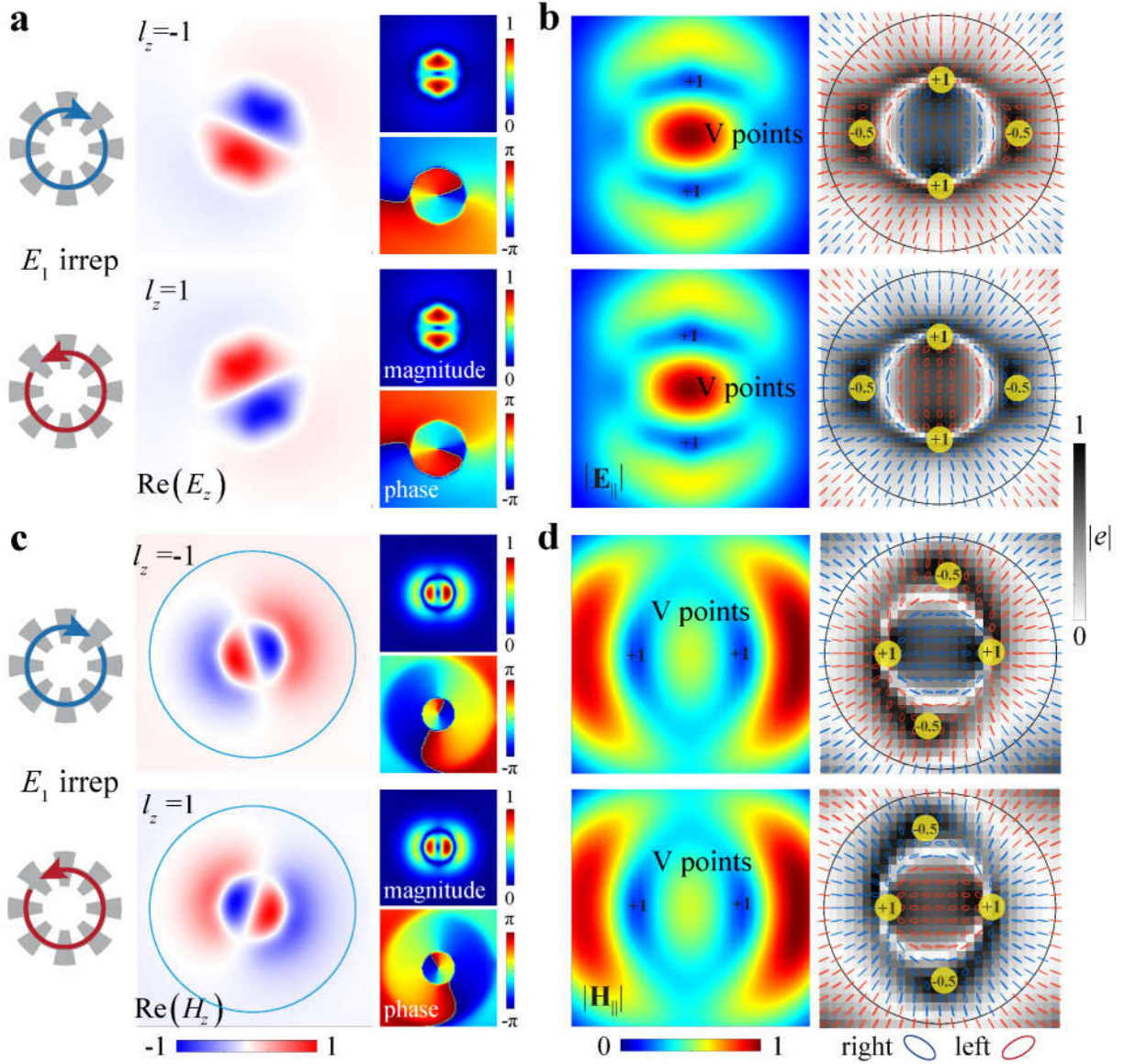

**Fig. 5. Phase and polarization singularities belonging to the two dimensions of the $E_1$ irrep. a,** The real part, magnitude and phase distributions of $E_z$. **b,** The magnitude and polarization distributions of the nest of polarization vortices in $\boldsymbol{E}_{\parallel}$, which is tiled by four polarization vortices whose topological charges are marked by four yellow circles, respectively. **c,** The real part, magnitude and phase distributions of $H_z$. **d,** the magnitude and polarization distributions of the nest of polarization vortices in $\boldsymbol{H}_{\parallel}$, which is tiled by four polarization vortices whose topological charges are marked by four yellow circles, respectively. In the figures, the circular region encircled by the black solid circles denotes the region at which the meron-antimeron pairs occur. In **b** and **d**, the V points with topological charges of +1 are marked in the magnitude plots of the in-plane fields (see the left column of **b** and **d**). As a remark, at the V points the magnitudes of the in-plane fields are zero. Furthermore, the absolute-value of the ellipticity distribution $e = b/a$ of the in-plane polarization states are plotted (see the right column of **b** and **d**), where $a$ and $b$ denote the semi-major and -minor axes of a polarization ellipse. At $|e| = 0$ and 1, the polarization states are linear and circular polarizations. Therefore, the inner white circles, which link all the V points and where $|e| = 0$, mark the L-line singularity maps of the nests of polarization vortices, which exactly define the boundaries of the spin merons and antimerons in Fig. 4.[5] In **c** and **d**, the blue and black circles are the white and black ones in **Fig. 3**, respectively. Also, the V points in **b** and **d** tend to split into two C points with the same topological charges of 0.5 and the opposite handedness (one left-handed and the other right-handed).

## Topology origins of the localized plasmonic topological quasiparticles

We further analyze lower dimensional optical singularities in plasmonic electric/magnetic fields, i.e., phase and polarization singularities, since the optical quasiparticles as higher dimensional singularities can be viewed as the synthesis of lower dimensional ones[9]. Fig. 5 depicts the phase and polarization singularities in out-of- and in-plane electric and magnetic fields, corresponding to the $E_1$ irrep. From Fig. 5a, it can be observed that the out-of-plane electric field $E_z$ carries a phase singularity at the center whose order is -1, thus exhibiting a field pattern of a scalar vortex with topological charge of -1. Accordingly, the real part of the component $\text{Re}(E_z)$ exhibits a dipolar pattern which revolves around the z axis and carries OAM of -1. From Fig. 5b, it can be observed that the in-plane electric field $\mathbf{E}_\parallel$ exhibits a nest of polarization vortices which is tiled by four polarization vortices. Two of them carry the V-type polarization singularities, i.e., the V points, with topological charges of 1; and the other two carry the C-type polarization singularities, i.e., the C points, with topological charges of -0.5. In addition, L-line polarization singularities are also observed in Fig. 5b and Fig. 5d, where the ellipticities are zero.

Here, we find a conservation law of total topological charges in the nest of polarization vortices. Previously, we demonstrated that each 2D irrep of the group under consideration corresponds to two degenerate polarization vortices with the same topological charge of 1 but with the opposite handedness in the in-plane electric fields[48]. Seemingly contradicting, here four polarization vortices with different topological charges appear, which seem to violate our previous results. Nevertheless, we must emphasize that irrep only regulates the total topology charge of a field distribution, i.e., the total topological charge of the four vortices is still 1, suggesting the total topological charge controlled by symmetries be conserved. Such a conservation law is strictly imposed by the symmetries of resonator and fundamentally governs the creation and annihilation of polarization singularities in the nest of polarization vortices.

The topology origins of the harmonic field meron-antimeron pairs in the $E_1$ irrep in Fig. 2a can be traced to the above scalar vortices with phase singularities and polarization vortex nests with polarization singularities. Take the meron-antimeron pair in the top panel of Fig. 2a as an example. Two arms of the dipolar pattern (see $\text{Re}(E_z)$ in Fig. 5a) contribute to the out-of-plane components of the meron and antimeron in the pair, and two polarization vortices with the V points in Fig. 5b contribute to the in-plane components. The core of the meron (antimeron) in the pair is up (down), resulting from the V point with the vanishing in-plane field intensity and the positive (negative) $\text{Re}(E_z)$. The topological charge of the scalar vortex in $E_z$ defines the OAM carried by the related meron-antimeron pair. The same story applies to the other meron-antimeron pairs in Fig. 2 and Fig. 3 (see more details in SI 3). It is worth noting that in Fig. 5c, the pattern of $\text{Re}(H_z)$ consists of two dipolar patterns viewed along the radial direction (see the region encircled by the blue circle in Fig. 5c). This suggests that $\text{Re}(H_z)$ flips twice along the radial direction in the region and thus gives birth to the target meron-antimeron pairs in Fig. 3a. The topology origins of steady spin (anti)merons in Fig. 4 can be explained with the framework of plasmonic SOI by considering the nontrivial field topologies of **E** and **H**.[5,45] The orientations of spin vectors **S** can be readily checked by their definition together with the vectorial distributions of **E** and **H** in Fig. 2 and Fig. 3. Note that the L-line singularity maps of the polarization vortex nest in magnetic fields (Fig. 5d and SI 3) define the boundaries of the spin merons or antimerons.[5] Owing to the symmetry protection of the scalar and polarization vortices, the above topological quasiparticles are strictly protected by the symmetries of resonator, to be specific, are originated from the 2D irreps of the groups formed by the symmetries.

## Experimental verification

To verify the numerical results, we fabricated a sample consisting of a resonator and a one-port feeding network (see Fig. 6a). The emission spectrum of the sample is measured and plotted in Fig. 6c, from which we observe that the measured spectrum shows good consistency with the numerical result in Fig. 1b. Only slight frequency discrepancies appear, mainly caused by the numerical and fabrication errors. To further confirm the existence of topological quasiparticles in doubly degenerate orbitals, we fabricated two samples consisting of the same resonator but with two different feeding networks (each having two angled and phased feeding lines, see Fig. 6b). The two feeding networks are built as the mirror of each other with respect to the $xoz$ plane. They are designed to excite two types of topological quasiparticles belonging to two dimensions of the $E_1$ irrep, including the harmonic meron-antimeron pairs in electric and magnetic fields in Fig. 2a and Fig. 3a, and the steady meron and antimeron in spin textures in Fig. 4a. We measured the $E_z$ components of the two topological quasiparticles (see details in Methods and SI 4), as illustrated in Fig. 6d, from which we observe that the measured real part, magnitude and phase distributions of $E_z$ agree well with the numerical results in Fig. 5a. Therefore, all experimental results demonstrate the validity of the theory and numerical results.

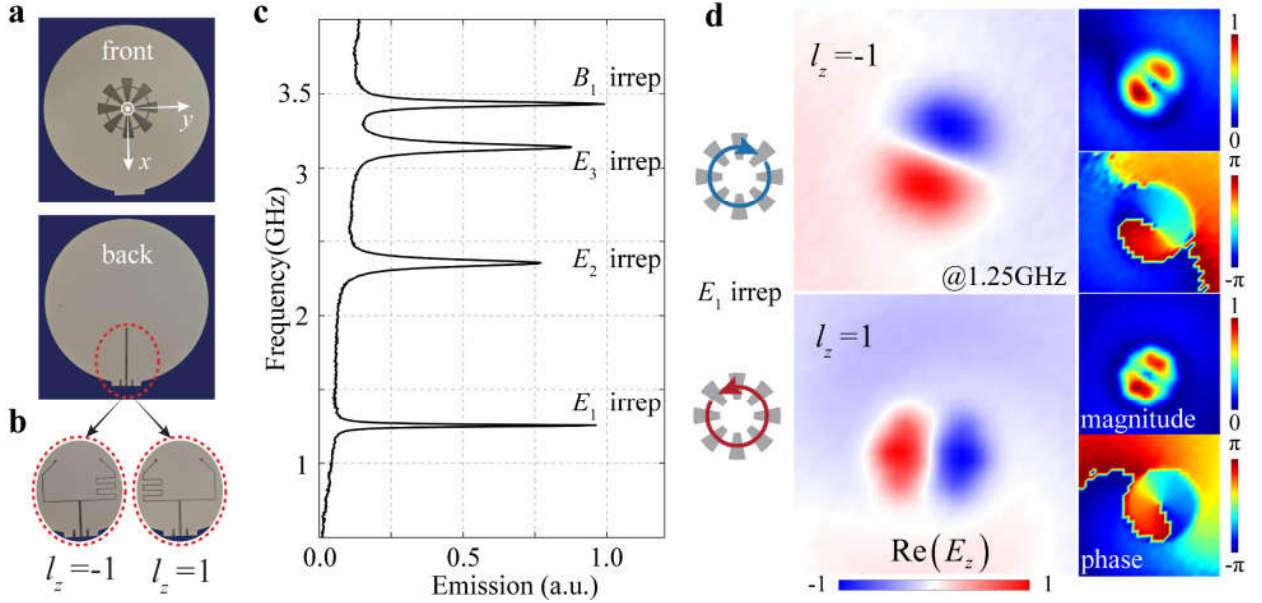

**Fig. 6. Experimental verification. a**, The first fabricated sample consisting of the resonator and one-port feeding network. **b**, The other two fabricated samples consisting of the same resonator but with two two-port feeding networks which are the mirror of each other with respect to the $xoz$ plane. Note that in **b** only the feeding networks of the two samples are presented. **c**, The emission spectrum of the one-port feeding sample, corresponding to Fig. 1b. **d**, The measured out-of-plane electric field components belonging to the two dimensions of the $E_1$ irrep, including their real part, magnitude and phase distributions in the cut plane of $z = 10mm$ above the resonator, corresponding to Fig. 5a. In **b**, the geometric angles $\theta$ and the dynamic phases $\psi$ of the two feeding networks are: $\theta = -\frac{\pi}{2}$ and $\psi = \frac{\pi}{2}$ for the left panel and $\theta = \frac{\pi}{2}$ and $\psi = \frac{\pi}{2}$ for the right panel, which are evaluated by Eq. (1) according to $l_z$. In **c**, the four excited eigenstates belonging to the four irreps occurred at 1.25, 2.36, 3.14, and 3.43 GHz.

## Conclusions

The above discussions are designed for the doubly degenerate irreps of $D_8$ group. For the nondegenerate $B_1$ irrep in Fig. 1b, it is demonstrated that the harmonic meron-antimeron pairs also exist in the plasmonic electric and magnetic fields (see Methods), whose absolute skyrmion numbers are 4 just as the irrep/orbital index of the $B_1$ irrep. However, since the elements of the $B_1$ irrep only consist of real numbers, i.e., -1 and +1 (see SI 2), the harmonic meron-antimeron pairs do not exhibit any vorticity and thus do not carry any OAMs. Apart from the harmonic pairs, we find that steady meron-antimeron pair also exists in the spin texture of the $B_1$ irrep (see methods), enriching the topological quasiparticle zoology in the rotational photonic systems.

It is important to point out that the scheme we demonstrate here is not restricted to plasmonic resonators with the $D_8$ group symmetries but generally applicable to the photonic resonators with any $D_M$ group symmetries (see SI 5 for a demonstration based on symmetry arguments). For the case of the $D_M$ group symmetries, our conclusions can be generalized to: 1) the number of the doubly degenerate irreps of $D_M$ group determines the absolute skyrmion numbers of meron-antimeron pairs carrying OAMs. Specifically, for even $M$, the allowed absolute skyrmion numbers range from 1 to $\left(\frac{M}{2}-1\right)$; and for odd $M$, range from 1 to $\left(\frac{M-1}{2}\right)$; 2) for a specific 2D irrep (e.g., the $j$-th 2D irrep), topological quasiparticles belonging to two dimensions of the irrep carry OAMs of $l_z = -j$ and $l_z = j$, which thus are degenerate orbitals; and 3) for the $j$-th 2D irrep, the two dimensions labelled by $l_z = -j$ and $l_z = j$ of the irrep always correspond to the isolated antimeron and meron with skyrmion numbers of -1/2 and 1/2, respectively, i.e., the chirality of the (anti)merons is locked with the parity of the irrep's dimensions or orbits. Such conclusions can also be transposed to photonic/plasmonic systems with arbitrary $D_M$ group symmetries owing to the generality of symmetry arguments, such as circular chromophore or nanopillar arrays[40,49], microring resonators[37], and plasmonic vortex lenses[39,50]. The above discussions offer a method to classify topological quasiparticles according to their symmetries (see Methods and Extended Data Table 1).

In summary, here we extend the isolated meron-antimeron pairs to photonic systems for the first time, which as new forms of the optical quasiparticles exist in the doubly degenerate orbitals of rotationally symmetric photonic systems. The isolated target meron-antimeron pairs are also exploited. Such (target) meron-antimeron pairs carry OAMs and exhibit deep-subwavelength features. We demonstrated that their absolute skyrmion numbers are strictly imposed by the orbital/irrep indices of the doubly degenerate irreps and thus protected by the symmetries of the resonator. In addition, we demonstrated the existence of isolated spin (anti)merons in the degenerate orbitals, which are normally believed to exist only in lattices or clusters in 2D ferromagnetic or photonic systems[5,18]. A chirality (of the (anti)meron) - parity (of the irrep) lock is further disclosed. The topological origins of the topological quasiparticles are revealed as well. Our findings shed light on the meron-antimeron pairs in plasmonics, which can greatly enrich the spectrum of topological quasiparticles in optics and can be readily transferred to other wave systems owing to the generality of the symmetry arguments.

## Methods

**Sample fabrications and measurements**

The layered configurations of three fabricated samples are illustrated in SI 1. The resonators and the feeding networks in Fig. 6a are made of copper with a thickness of 0.034 mm whose surfaces are electroplated with tin. The metallic resonators and feeding networks are printed on the dielectric plates (Rogers 4350B with a relative permittivity of 3.3). The samples are fabricated with printed circuit board (PCB) manufacturing technology. The samples were measured by a near-electric-field scanning system in an anechoic chamber (see details in SI 4). The scanning system consists of a servo actuator, a coaxial near-field probe, a vector network analyzer (VNA) and connection cables. A 50 Ω SMA (Sub-Miniature version A) connector was welded onto the microstrip lines of the feeding networks to receive the input signal from the VNA. In the fabrication, glue was used to bond the resonator and the feeding system together, which can lead to discrepancies between simulations and experiments.

**Meron-antimeron pairs in the $B_1$ irrep and their topological origins**

From Fig. 1b, the eigenstate belonging to the $B_1$ irrep can be excited by a single point source. The electric and magnetic components, and spin texture of the excited eigen spoof LSP state are illustrated in Extended Data Fig. 1-Extended Data Fig. 3, respectively. Surprisingly, topological (target) meron-antimeron pairs appear in electric, magnetic and spin fields of spoof LSP state.

**Harmonic isolated meron-antimeron pair in the electric field.** From Extended Data Fig. 1, it can be observed that there exist four pairs of Néel-type merons and antimerons in the electric field of the spoof LSP state, i.e., an isolated meron-antimeron pair with absolute skyrmion number of 4. Extended Data Fig. 1b depicts the $z$ component of the pair, from which it can be observed that an octupole state appears, which can be viewed as superposition of two scalar vortices with topological charges of $\pm 4$.[42] The pair does not exhibit any vorticity and carries no OAM since elements of the $B_1$ irrep jump from -1 to 1(see Table1 in SI 3 and the phase distribution in Extended Data Fig. 1b). Extended Data Fig. 1c depicts the in-plane components of the pair, from which it can be observed that a nest of polarization vortices with eight V points carrying a topological charge of 1 occurs. The cores of the merons (antimerons) in the pair are up (down), resulting from the V point with vanishing in-plane field intensity and the positive (negative) $\text{Re}(E_z)$. The pair is harmonic and highly localized.

**Harmonic isolated target meron-antimeron pair in the magnetic field.** From Extended Data Fig. 2a, it can be observed that 1) a meron-antimeron pair with an absolute skyrmion number of 4 appears within the circular region encircled by a black solid circle; and 2) the pair is surrounded by another meron-antimeron pair with an absolute skyrmion number of 4 in the annular region delineated by the black and white solid circles. Therefore, in the region encircled by the solid black circle, the magnetic field vectors flip an angle of $2\pi$ as going along the radial direction and thus the meron-antimeron pair in Extended Data Fig. 2 is a target one. Extended Data Fig. 2b depicts the z component of the pair, from which it can be observed that an octupole state appears. Also, Extended Data Fig. 2b suggests that $\text{Re}(H_z)$ flips twice along the radial direction in the region and thus gives rise to the target meron-antimeron pair. The pair does not exhibit any vorticity and carries no OAM since elements of the $B_1$ irrep jump from -1 to 1(see Table 1 in SI 2 and the phase distribution in Extended Data Fig. 2b). Extended Data Fig. 2c depicts the in-plane components of the inner meron-antimeron pair, from which it can be observed that a nest of polarization vortices with eight V points carrying topological charges of 1, occurs. The cores of the merons (antimerons) in the pair are up (down), resulting from the V point with vanishing in-plane field intensity and the positive (negative) $\text{Re}(H_z)$. The pair is harmonic and highly localized.

**Steady isolated meron-antimeron pair.** Different from the harmonic meron-antimeron pairs in plasmonic electric/magnetic fields, we find that a steady isolated meron-antimeron pair emerges from plasmonic spin textures (see Extended Data Fig. 3), which is also different from the steady isolated (anti)merons in the doubly degenerate irreps. The pair carries the absolute skyrmion number of 1. The vectorial configuration of the spin meron-antimeron pair can be explained within the framework of plasmonic SOI by considering the nontrivial field topologies of **E** and **H**.[5,45] The orientations of spin vectors **S** can be readily checked by their definition in the main text together with the vectorial distributions of **E** and **H** in Extended Data Fig. 1 and Extended Data Fig. 2. It is worthing noting that the above discussions about the $B_1$ irrep is also applicable to the case of the $B_2$ irrep in Table S1 in SI 2.

**Extended analysis to upper bound of absolute skyrmion numbers**

The absolute skyrmion numbers as topological invariants of the meron-antimeron pairs are rigorously dictated and upper-bounded by the orbital indices of doubly degenerate irreps of the underlying symmetry groups. Here we consider a generic rosette group $D_M$, where $M$ is an arbitrary integer[38]. Such a group defines $\lfloor \frac{M-1}{2} \rfloor$ doubly degenerate irreps (also see SI 5)[33]. Accordingly, an upper bound is introduced for the orbital index of doubly degenerate orbitals. According to Eq. (2), we can reveal the upper bound of the absolute skyrmion numbers carried by the field meron-antimeron pairs in the doubly degenerate irreps,

$$N \leq \left\lfloor \frac{M-1}{2} \right\rfloor. \tag{4}$$

Namely, $N$ ranges from 1 to $\lfloor \frac{M-1}{2} \rfloor$. Such the conclusion also applies to a generic cyclic group $C_M$ whose doubly degeneracies result from time reversal symmetry[33].

**Topological quasiparticles as irreps of rosette groups: symmetry classifications**

According to the above discussions, we can enumerate all topological quasiparticles belonging to the irreps of a generic rosette group $D_M$, as shown in Extended Data Table 2. From the table it can be observed that 1) the skyrmions with skyrmion number of 1 belonging to the irrep $A_1$; 2) the nondegenerate meron-antimeron pairs belong to the $A_2$, $B_1$ and $B_2$ irreps, which exhibit no vorticity and thus carry no OAM; and 3) the doubly degenerate meron-antimeron pairs belong to the $E_h$ irreps ($h$ ranges from 1 to $\lfloor \frac{M-1}{2} \rfloor$ according to Eq. (4)), which exhibit vorticity and thus carry OAM. The table gives all categories of topological quasiparticles in photonic/plasmonic systems with rotational symmetries. As basic blocks they can be further generalized to the forms of lattices[6,45]. Note that in the table, $M$ is even. When $M$ is odd, the $B_1$ and $B_2$ irreps will vanish. When reflection symmetries vanish, the $D_M$ group degrades into the $C_M$ group. Correspondingly, the $A_2$ and $B_2$ irreps will vanish for even $M$ case, and for odd $M$ case, the $A_2$, $B_1$ and $B_2$ irreps will vanish[33]. Vanishing irreps lead to that the related topological quasiparticles in the table will not be further supported by photonic or plasmonic systems.

It is worth mentioning that the statement that a topological quasiparticle belongs to an irrep refers to that the vector configuration (or vectorial function distribution) of the topological quasiparticle belongs to the irrep. Such the vectorial function is not limited to the electric or magnetic fields, which should be understood as a generic vectorial function. Such generic vector functions are termed basis functions of the irreps, as they transform according to the rules specified by the irreps. We only identify them as the electric or magnetic fields in our case owing to the commutation relations between group projection operators and the impedance operators of photonic/plasmonic systems[38]. They can also be identified as other physical fields in other physical systems. For example, considering the commutation relations between group projection operators and the Hamiltonian of a quantum system, such the function can be identified as the wave functions of electrons[33].

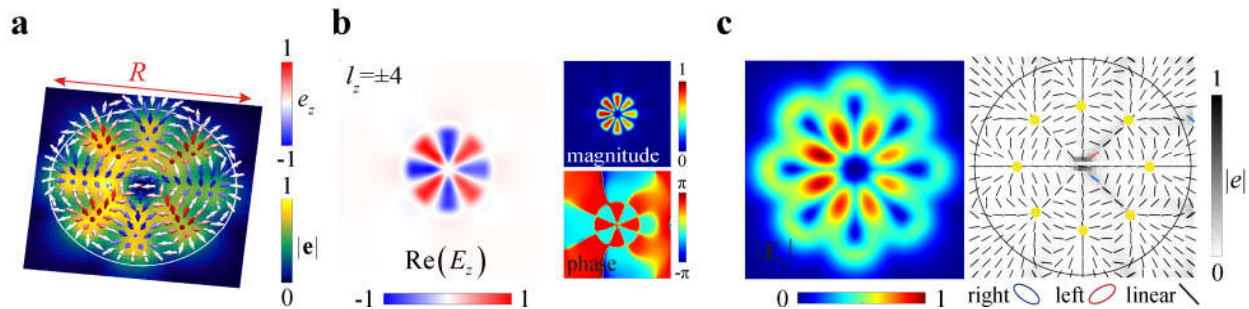

**Extended Data Fig. 1. Harmonic meron-antimeron pairs in plasmonic electric field belonging to the $B_1$ irrep. a,** the excited meron-antimeron pair with an absolute skyrmion number of 4 but carrying no OAM at 3.33 GHz. $|e|$ denotes the normalized distributions of electric fields, and $e_z$ denotes the z component of normalized electric fields. The colors of the vectors are coded from blue to red to denote the latitude angle $e_z$ varying from -1 to 1. **b,** the real part, magnitude and phase distributions of $E_z$. **c,** the magnitude and polarization distributions of the nest of polarization vortices in the $E_\parallel$. In the polarization distribution of **c**, eight V-point singularities are marked by yellow dots and their topological charges are 1. In **a**, $R = 43\ mm$, and the meron-antimeron pair is on the subwavelength scale and is down to $\lambda/2.1$, indicating that each meron/antimeron in the pair can be as small as $\lambda/16.8$.

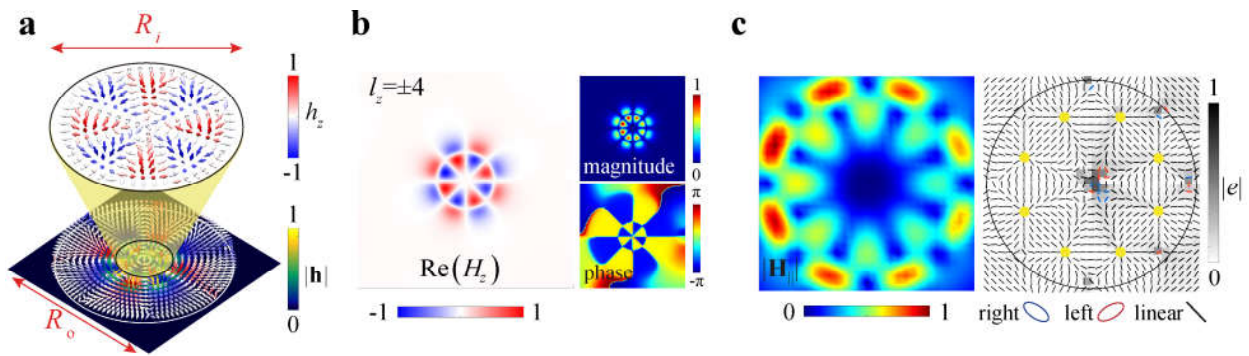

**Extended Data Fig. 2. Harmonic target meron-antimeron pairs in plasmonic magnetic field belonging to the $B_1$ irrep. a,** the excited target meron-antimeron pair with absolute skyrmion number of 4 but carrying no OAM at 3.33 GHz. $|h|$ denotes the normalized distribution of magnetic fields, and $h_z$ denotes the z component of the normalized fields. The colors of the vectors are coded from blue to red to denote the latitude angle $h_z$ varying from -1 to 1. **b,** the real part, magnitude and phase distributions of $H_z$. **c,** the magnitude and polarization distributions of the nest of polarization vortices in the $H_\parallel$. In the polarization distribution of **c**, eight V-point singularities are marked by yellow dots, whose topological charges are 1. In **a**, the inner and outer diameters of the circles denoted by black and white colors are $R_i = 28\ mm$ and $R_o = 86\ mm$, respectively; and the meron-antimeron pair in the inner circular circle is on the subwavelength scale and can be down to $\lambda/3.2$, indicating that the size of each meron/antimeron can be as small as $\lambda/25.7$.

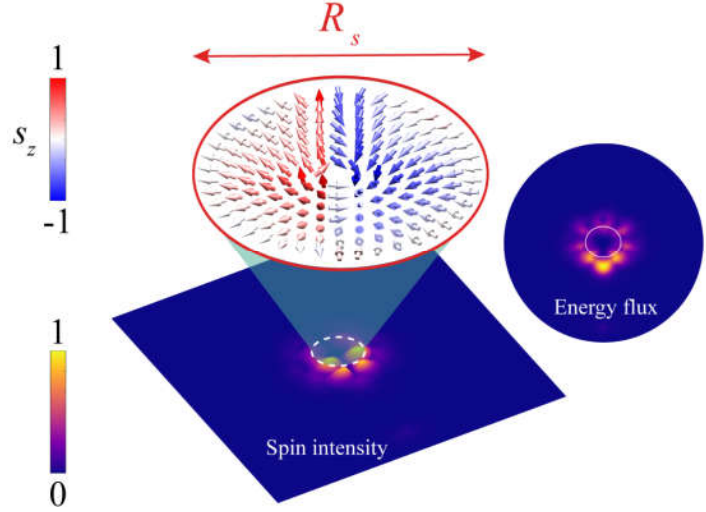

**Extended Data Fig. 3. Steady meron-antimeron pair in plasmonic spin texture belonging to the $B_1$ irrep.** The colors of the vectors are coded from blue to red to denote the latitude angle of $s_z$ varying from -1 to 1. The diameter of the red solid circles is $R_s = 19.2 \ mm$. In the inset, the magnitude distribution of energy flux $\boldsymbol{p} = \frac{1}{2}Re(\boldsymbol{E} \times \boldsymbol{H}^*)$ is plotted, from which it can be observed that the energy is highly localized within a subwavelength regime. The size of the meron and antimeron in the pair is as small as $\lambda/9.4$.

**Extended Data Table 2. Topological quasiparticles as the basis functions of irreps of rosette groups $D_M$.** Note that the skyrmions belonging to the $A_1$ irrep carry skyrmion number of 1 instead of absolute skyrmion number.

| $D_M$ | Irrep index | Topologial quasiparticels | Vorticity (OAM numbers) | (absolute) Skyrmion numbers |
|---|---|---|---|---|
| $A_1$ | $j=0$ | skyrmions[46] | No ($l_z = 0$) | 1 |
| $A_2$ | $j=0$ | Meron-antimeron pairs | No ($l_z = \pm 2M$) | $N = |l_z| = 2M$ |
| $B_1$ | $j=\frac{M}{2}$ | Meron-antimeron pairs | No ($l_z = \pm M$) | $N = |l_z| = M$ |
| $B_2$ | $j=\frac{M}{2}$ | Meron-antimeron pairs | No ($l_z = \pm M$) | $N = |l_z| = M$ |
| $E_h \left( h \in \{1, ..., \lfloor \frac{M-1}{2} \rfloor \} \right)$ | $j=h$ | Meron-antimeron pairs | Yes ($l_z = -h$) | $N = |l_z| = h$ |
|  | $j=M-h$ | Meron-antimeron pairs | Yes ($l_z = +h$) | $N = |l_z| = h$ |